\documentclass[a4paper,fleqn,usenatbib]{mnras}
\usepackage{natbib}

\usepackage[T1]{fontenc}
\usepackage{ae,aecompl}

\usepackage{graphicx}
\usepackage{color}
\usepackage{amssymb}
\usepackage{amsmath}
\usepackage{multicol}
\usepackage{multirow}
\input{colordvi.tex}

\title{Analysing the 21cm signal from the Epoch of Reionization with artificial neural networks}
\author[H.Shimabukuro et al.]
  {Hayato Shimabukuro$^1$ and Benoit Semelin$^1$ \\ 
  $^1$Sorbonne Universit\'es, UPMC, LERMA, Observatoire de Paris, PSL research university, CNRS, F-75014, Paris, France
   }

\begin{document}

\label{firstpage}
\pagerange{\pageref{firstpage}--\pageref{lastpage}}
\maketitle

\begin{abstract}
The 21 cm signal from the Epoch of Reionization should be observed within the next decade. While a simple statistical detection is expected with SKA pathfinders, the SKA will hopefully produce a full 3D mapping of the signal. To extract from the observed data constraints on the parameters describing the underlying astrophysical processes, inversion methods must be developed. For example, the Markov Chain Monte Carlo method has been successfully applied. Here we test another possible inversion method: artificial neural networks (ANN). We produce a training set which consists of 70 individual sample. Each sample is made of the 21 cm power spectrum at different redshifts produced with the 21cmFast code plus the value of three parameters used in the semi-numerical simulations that describe astrophysical processes. Using this set we train the network to minimize the error between the parameter values it produces as an output and the true values. We explore the impact of the architecture of the network on the quality of the training. Then we test the trained network on the new set of 54 test samples with different values of the parameters. We find that the quality of the parameter reconstruction depends on the sensitivity of the power spectrum to the different parameters at a given redshift, that including thermal noise and sample variance decreases the quality of the reconstruction and that using the power spectrum at several redshifts as an input to the ANN improves the quality of the reconstruction. We conclude that ANNs are a viable inversion method whose main strength is that they require a sparse exploration of the parameter space and thus should be usable with full numerical simulations.

\end{abstract}


\section{Introduction}

According to the standard hierarchical structure formation theory based on the $\Lambda$CDM cosmology model, density perturbations grow into small haloes where the first stars and galaxies form \citep[e.g.][]{2002Sci...295...93A, 2002ApJ...564...23B,yos,2013RPPh...76k2901B}.  These astrophysical sources produce ultraviolet~(UV) ionizing photons that escape into the intergalactic medium~(IGM) which is eventually reionized.  The epoch when this process occurs is called the ``Epoch of Reionization''~(EoR).  The detailed astrophysics of the EoR is currently poorly understood because no observation has yet probed the early and middle stages of the EoR although current observations have provided fruitful information on late stage of the EoR.  For example, the absorption spectra of high-$z$ quasars indicate that reionization was complete by $z\sim 6$ \citep[e.g.][]{Fan:2006dp} and the number density of Ly$\alpha$ emitter galaxies at $z=6-7$ implies that the neutral hydrogen fraction increases at $z>6$ \citep[e.g.][]{2014ApJ...797...16K}.  On cosmological scales, reionization induces Thomson scattering of CMB photons off free electrons.  The optical depth of Thomson scattering measured by Planck is $\tau_{e}=0.058\pm0.012$ which corresponds, using an instantaneous reionization toy model, to a reionization at $z=8.8\pm0.9$~\cite{2016arXiv160503507P}. 

To further improve our understanding of the EoR, observations targeted on the cosmological redshifted 21cm signal from the EoR are on-going.  The 21cm signal emission is due to the hyperfine structure of neutral hydrogen atoms and is expected to be a powerful tool to probe the neutral IGM, yielding both astrophysical and cosmological information such as matter density fluctuations and the ionization state and thermal history of the IGM at high redshift~\cite{fur, 2012RPPh...75h6901P}.   

Recently, some first-generation radio interferometers have been attempting to detect statistically the 21cm signal from the EoR, such as the Murchison Wide field Array (MWA) ~\cite{Tingay:2012ps}, the LOw Frequency ARray (LOFAR)~\cite{Rottgering:2003jh} and the Precision Array for Probing the Epoch of Reionization (PAPER)~\cite{2015ApJ...801...51J}.  These observational efforts have resulted in upper limits of the 21cm power spectrum and on the ionized state of the IGM at $z\sim 8$~ \citep[e.g.][]{2014ApJ...788..106P, 2015ApJ...809...61A, 2015ApJ...809...62P, 2016MNRAS.455.4295G, 2016ApJ...833..102B}. Furthermore, future instruments such as the Square Kilometre Array (SKA)~\cite{2013ExA....36..235M} and Hydrogen Epoch of Reionization Array (HERA)~\cite{2016arXiv160607473D} are designed to detect the 21cm signal power spectrum with higher signal to noise ratio and at higher redshift, during the Cosmic Dawn. The SKA should also be able to image the signal in 3D, which requires high sensitivity  on small enough scale (that is sufficient collecting area in a large enough core).

So we expect a wealth of 21cm data in the near future. Then we face the fundamental question: what we can learn from the data?  From theoretical and numerical works, we have already some insights on the process of reionization.  For instance, considering galaxies in relatively massive host haloes results in larger and more uniform ionized bubbles and imprints a peak at larger scales in the 21cm power spectrum \citep[e.g.][]{2007MNRAS.377.1043M,2012MNRAS.423.2222I}.  On the other hand, abundant and small minihaloes, which serve as absorption systems by self-shielding from ionizing photons, result in small and disjointed ionized regions~\citep[e.g.][]{2006MNRAS.366..689C}. To extract information on the EoR from the observed 21cm signal, we need to be able compute the 21cm signal from the basic physics of reionization \citep[e.g.][]{2015aska.confE..13S}. The most self-consistent method is to run numerical simulations. There are currently two approaches.  The first is to implement the full radiation hydrodynamics (RHD).  This type of simulations is relevant if small scales, where feedback effects such as photo-heating play an important role, are resolved \citep[e.g.][]{2000MNRAS.314..611C, 2002ApJ...575...49R,2013MNRAS.428..154H}.  The second approach is to run large scale simulations where radiative transfer~(RT) is computed in post-processing~\citep[e.g.][]{2006MNRAS.369.1625I, 2007ApJ...671....1T,2014MNRAS.439..725I}.  This type of simulations are able to account for large scales fluctuations in the signal but they require sub-grid modelling to treat processes on galaxy scales (feedback, ionizing photons escape fraction, etc.).

Although numerical study based on simulations is the most consistent method, it implies a large computational cost. This disadvantage is somewhat alleviated by using a semi-numerical approach instead of RHD or RT simulations. In most semi-numerical simulations, the production of ionizing photons is calculated based on the excursion set formalism, using analytically derived halo mass functions, and density fluctuations are obtained with linear perturbation theory~\cite{Mesinger:2007pd,2010MNRAS.406.2421S,2011MNRAS.411..955M}.   However, the growth of HII regions is simply evaluated by considering the balance between the production of ionizing photons and the number of neutral hydrogen atoms \cite{2004ApJ...613....1F}.  Recently, the effect of recombination is also taken into account \citep[e.g.][]{2014MNRAS.440.1662S}. The results produced by the semi-numerical approach shows good agreement with those by RT simulations on large scale ($\gtrsim$ Mpc)~\cite{2007ApJ...654...12Z, 2011MNRAS.414..727Z, 2011MNRAS.411..955M}. 
  
To maximize the scientific return of the upcoming observations it is important to establish systematic procedures to derive constraints on the EoR modeling parameters from the observed 21-cm data. Recently, several works have studied how such constraints can be obtained by exploring the EoR parameter space with semi-numerical simulations.  For example, Fisher analysis~\cite{2014ApJ...782...66P,2016PASJ...68...61K, 2016arXiv160800372S,2016MNRAS.457.1864L} and Bayesian parameter inference such as the Markov Chain Monte Carlo~(MCMC) approach~\cite{2015MNRAS.449.4246G,2016MNRAS.461.2847B} have been applied.

In this work, we suggest a new approach for parameter reconstruction based on a machine learning method. Machine learning is one of the hot topics in data science as a method to deal with big data. It is currently applied to many fields such as pattern recognition or search engine. The main purpose of machine learning is to find approximate functions that, given the input produce the desired outputs. This is achieved by ``learning" from training datasets with known inputs and outputs \citep[e.g.][]{2010IJMPD..19.1049B, 2012amld.book...89B}. 

Recently, machine learning methods have been applied in the field of astronomy.   For example, learning from huge galaxy image catalogs helps with morphological classification of galaxies~\cite{1996MNRAS.283..651F,1996MNRAS.283..207L,2010MNRAS.406..342B}. Machine learning  can also help with the selection and classification of transients ~\cite{2016PASJ...68..104M,2016arXiv160907259M}.  Applying machine learning to a large sample of spectroscopic and photometric galaxies data can improve the accuracy of estimates of photometric redshifts~\cite{2004PASP..116..345C, 2004A&A...423..761V,2017NewA...51..169S}.  Using simulated gravitational wave data with noise as learning sample, machine learning can help us search for gravitational wave signals from noisy real data~\cite{2015CQGra..32x5002K}.  In the context of cosmology, machine learning is used to model galaxy formation~\cite{2016MNRAS.455..642K,2016MNRAS.457.1162K} or to make templates of nonlinear matter power spectrum~\cite{2012MNRAS.424.1409A,2014MNRAS.439.2102A}. Closer to our field of interest, there is a study applying a machine learning method to estimate the escape fraction of ionizing photons during the EoR~\cite{2016ApJ...827....5J}.  In this study, they show how machine learning can estimate the Lyman continuum escape fraction by using mock spectroscopic simulation data.  In our work, using a simple astrophysical parameterization of the EoR, we apply artificial neural network (ANN), which is one of the machine learning methods, to reconstruct the parameter values from the 21cm power spectrum data.

This paper is organised as follows.  In section \ref{sec:21cm}, we introduce the cosmological 21cm signal and the EoR parameterization we focus on.  In section \ref{sec:ANN}, we describe our artificial neural network and test the impact of its chosen architecture.  In section \ref{sec:result} we show our main results, and we give a summary and discussion in section~\ref{sec:summary}.  Throughout this paper, we employ the best fit cosmological parameters obtained by \cite{2016A&A...594A..13P}.  

\section{Cosmological 21cm signal}\label{sec:21cm}
\subsection{Introduction of 21cm power spectrum}
\label{sec:21cmps}
The brightness temperature  for the $21$ cm signal is given by \citep[e.g.][]{2013ExA....36..235M}:

\begin{align}
\delta T_{b}(\nu) &= \frac{T_{{\rm S}}-T_{\gamma}}{1+z}(1-e^{-\tau_{\nu_{0}}})  \nonumber \\
                  &\quad \sim 27x_{{\rm H}}(1+\delta_{m})\bigg(\frac{H}{dv_{r}/dr+H}\bigg)\bigg(1-\frac{T_{\gamma}}{T_{{\rm S}}}\bigg) \nonumber \\
                  &\quad \times \bigg(\frac{1+z}{10}\frac{0.15}{\Omega_{m}h^{2}}\bigg)^{1/2}\bigg(\frac{\Omega_{b}h^{2}}{0.023}\bigg)\bigg(\frac{\Omega_b h}{0.031}\bigg) [{\rm mK}].
\label{eq:brightness}
\end{align}
Here, $T_{\rm S}$ and $T_{\gamma}$ represent the local spin temperature of the IGM and the CMB temperature, respectively.  $\tau_{\nu_{0}}$ is the local optical depth in the 21cm rest frame frequency $\nu_{0} = 1420.4~{\rm MHz}$, $x_{\rm H}$ is the local neutral fraction of the hydrogen gas, $\delta_{m}({\bold x},z) \equiv \rho/\bar{\rho} -1$ is the evolved matter overdensity,  $dv_{r}/dr$ is the local gradient of the gas velocity along the line of sight and $H(z)$ is the Hubble parameter.  All quantities are evaluated at redshift $z = \nu_{0}/\nu - 1$.  As we can see, the 21cm signal includes both astrophysical and cosmological information.  Thus, we can hope to use the 21cm signal to disentangle and quantify them \cite{2015aska.confE..12P}.

Let us now introduce the power spectrum of the 21cm fluctuations. We define the 21cm power spectrum as 


\begin{equation}
\langle \delta T_b({\bold k}) \delta T_b({\bold k^{'}})\rangle
= (2\pi)^3 \delta({\bold k}+{\bold k^{'}}) P_{21}({\bold k}).
\label{eq:ps_def}
\end{equation}
In our context, we use the {\it dimensional} 21cm power spectrum,  $k^{3}P(k)/2\pi^{2}$.
The 21cm PS would describe the statistical properties of the 21cm fluctuations perfectly if they were a gaussian random field.  However, the 21cm fluctuations are expected to deviate from a gaussian behaviour because of astrophysical effects such as ionization and X-ray heating.  Thus it is useful to compute higher order statistics and one-point statistics such as the bispectrum, the variance and the skewness to estimate non-gaussian features in the 21cm fluctuations~\citep[e.g.][]{2008MNRAS.384.1069B, 2009MNRAS.393.1449H, 2010A&A...523A...4B, 2014MNRAS.443.1113P, 2015MNRAS.451.4986S,2016MNRAS.458.3003S}. 

In order to generate the 21cm PS for a given set of astrophysical parameters, we use the publicly available code {\bf 21cmFAST}~\cite{2011MNRAS.411..955M}.  This code is based on a semi-numerical model of cosmic reionization and thermal history of the IGM.  It quickly generates maps and PS of the brightness temperature, matter density, velocity, spin temperature and ionization fraction at designated redshifts.

We performed simulations in a $200$ ${\rm Mpc}^{3}$ comoving box with $300^{3}$ grid cells for a wide range of EoR parameters described in the next section.  In our calculation, we use the 21cm power spectrum in the range $0.06 ~\rm Mpc^{-1}\le {\it k} \le 1.4 ~\rm Mpc^{-1}$ divided into 14 bins.

\subsection{EoR model parameters}

It is common to characterize EoR models with parameters and then examine the effect of changing the parameters on the 21cm signal.  We employ three key parameters  which are often used.  Let us briefly define these three parameters:\\

1. $\zeta$, {\it the ionizing} {\it efficiency}: $\zeta$ is the combination of several parameters related to ionizing photons escaping from high redshift galaxies and is defined as $\zeta=f_{\rm esc}f_{*}N_{\gamma}/(1+\overline{n}_{\rm rec})$ \cite{2004ApJ...613....1F, fur}.  Here, $f_{\rm esc}$ is the fraction of ionizing photons escaping from galaxies into the IGM and $f_{*}$ is the fraction of baryons locked into stars.  These parameters are extremely uncertain at high redshift~\cite{2008ApJ...672..765G,2009ApJ...693..984W}.  $N_{\gamma}$ is the number of ionizing photons produced per baryon in stars and $\overline{n}_{\rm rec}$ is the mean recombination rate per baryon. In our calculation, we explore a range of $10 \le \zeta \le 60$.  \\

2. $T_{{\rm vir}}$, {\it the minimum} {\it virial} {\it temperature} {\it of} {\it haloes} {\it producing} {\it ionizing} {\it photons}:  $T_{{\rm vir}}$ parameterizes the minimum mass of haloes producing ionizing photons during the EoR.  Typically, $T_{\rm vir}$ is chosen to be $10^{4} {\rm K}$, corresponding to the temperature above which atomic cooling becomes effective.  $T_{\rm vir}$ parameterizes the physics of star formation in high redshift galaxies. In haloes with 
virial temperature $> 10^{4}{\rm K}$ atomic cooling is sufficient to trigger gravothermal instability and thus star formation.  However, star formation is quenched if AGN or Supernovae feedback is effective and the IGM is heated up.  This leads to effective minimum virial temperature larger than $10^{4} $[\rm K].  In haloes with viral temperature $<10^{4}{\rm K}$, hydrogen molecule cooling is necessary. However, if stars begin to form in a halo, radiative feedback such as the photodissociation of $\rm H_{2}$ by Lyman-Werner photons may become effective and prevent the gas from cooling \cite{2013MNRAS.432.3340S, 2013MNRAS.432.2909F}.  Conversely, positive feedback, such as the enhancement of $\rm H_{2}$ molecules formation due to an increase in the free electrons density, tends to push the minimum virial temperature to lower value because cooling becomes more effective.  Thus $T_{\rm vir}$ parameterizes the uncertainty in the efficiency of radiative feedback. In our work, we explore $T_{\rm vir}$ ranging from $10^{3} {\rm K}$ to $5\times 10^{5} {\rm K}$.\\

3. $R_{\rm mfp}$, {\it the mean free path of ionizing photons} : The propagation of ionizing photons through the ionized IGM strongly depends on the presence of absorption systems and the sizes of ionized regions are determined by the balance between sinks and sources of ionizing photons \citep[e.g.][]{2011ApJ...743...82M}.  This process is modelled  by the maximum mean free path of ionizing photons, $R_{\rm mfp}$\cite{2014MNRAS.440.1662S}.  Physically, the mean free path of ionizing photons corresponds to the typical distance traveled by photons within ionized regions before they are abosorbed and is determined by the number density and the optical depth of Lyman-limit systems.  In our calculation, we explore $R_{\rm mfp}$ from 10 \rm Mpc to 60 \rm Mpc.

\section{Artificial neural networks}\label{sec:ANN}
\subsection{Architecture of ANN}
\begin{figure}
\includegraphics[width=1.0\hsize]{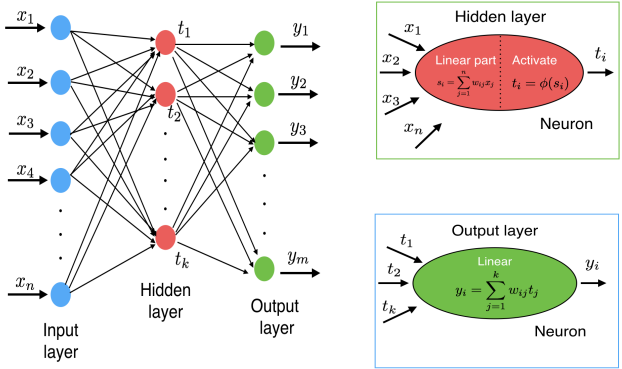}
\caption{Typical architecture of an artificial neural network.  The architecture of the ANN consists of an input layer, a hidden and an output layer of neurons. Each neuron connects the neurons in the next layer.}
\label{fig:fig1}
\end{figure}

In this section, we introduce artificial neural networks (ANN). ANNs are one of the machine learning methods and are a mathematical model inspired by the natural neuron network in our brain.  The main purpose of ANNs is to construct approximate functions which associate input data with output data.  In order to construct such a function, the ANN has to learn from `` {\it training data}''. The architecture of a simple class of ANN consists of three layers: the input layer, the hidden layer and the output layer.  Each of them has a number of neurons as shown in Fig.\ref{fig:fig1}.  In a more general case, we could choose the number of hidden layers and the number of neurons at each layer arbitrarily.   In our study,  we use 1 hidden layer.  Note that it is mathematically proven that neural networks with only 1 hidden layer can approximate any function with any accuracy if we use a large enough number of neurons \cite{Cybenko, Hornik}.

Let us briefly describe the architecture of our ANN. The input data  $x_{j}$ is fed to the $j$-th neurons in the input layer.  Each neuron in the input layer is connected to the $i$-th neuron in the hidden layer and a weight $w^{(1)}_{ij}$ is associated with the connection. The input to the $i$-th neuron in the hidden layer $s_{i}$ is a linear combination of all the input neurons with weight $w^{(1)}_{ij}$ :

\begin{equation}
s_{i}=\sum_{j=1}^{n} w^{(1)}_{ij}x_{j}.
\label{eq:hidden}
\end{equation}

Here, $n$ is the number of input data.  In the hidden layer, the $i$-th neuron is activated by an activation function $\phi$ such as its output is $t_{i}=\phi(s_{i})$.  Generally, the activation function is a nonlinear function. We use the sigmoid function $\phi(s)=1/(1+e^{-s})$.  The properties of the sigmoid function are such that it saturates and returns a constant output when the absolute value of the input is large and that it is a smooth and differentiable function.  Thanks to the nonlinear activation function, a trained ANN can express any function.

In the output layer, we compute linear combinations of the activated outputs of the neurons in the hidden layer with weights $w^{(2)}_{ij}$ and obtain the output vector:

\begin{equation}
y_{i}=\sum_{j}^{k}w^{(2)}_{ij}t_{j}
\label{eq:output}
\end{equation}
Here $k$ is the number of neurons in the hidden layer.  Note that we do not activate the output value.  The aim of training the ANN is to find a set of weights that ensures the output vectors produced by the ANN for a set of input vectors is sufficiently close to the desired output vectors.  Once we adjust the weights to reach this goal on a training sample, we can make predictions for output vectors for arbitrary input vectors outside of the training sample (for example, new observational data).

A popular algorithm to compute the trained weights is the ``{\it Back propagation algorithm}'' \cite{1986Natur.323..533R}.  We will describe this algorithm briefly in the following section.

\subsection{Back propagation algorithm}
In this section, we present the back propagation algorithm for the 1 hidden layer case.  We also show the back propagation algorithm for multiple hidden layers case in Appendix.

In order to quantify how well the output obtained by the ANN approximates the desired output for the training data set, we define the (total) cost function as:

\begin{equation}
E=\sum_{n=1}^{\it N_{\rm train}}E_{n}=\sum_{n=1}^{\it N_{\rm train}}\left[\frac{1}{2}\sum_{i=1}^{m}(y_{i,n}-d_{i,n})^{2}\right],
\label{eq:cost}
\end{equation}
where $N_{\rm train}$ is the number of training input vectors and $m$ is the number of neurons at the output layer. $y$ and $d$ are outputs of the ANN and the (desired) training output data, respectively. Our purpose is to find the weight set that minimises the cost function.  In order to find this weights set, we need to compute the partial derivative of $E$ with respect to the individual weights $w^{(l)}_{ij} (l=1,2)$ and find the local minimum of $E$ using gradient descent.

The weights are updated by gradient descent following the formula: 

\begin{equation}
\Delta w^{(l)}_{ij}=-\eta \frac{\partial E}{\partial w^{(l)}_{ij}}=-\eta \sum_{n=1}^{\it N_{\rm train}} \frac{\partial E_{n}}{\partial w^{(l)}_{ij}} 
\label{eq:gradient}
\end{equation}
Here, $\eta$ is a learning coefficient which controls how fast the weights are updated.  We used $\eta=0.01$.  We only need to calculate the derivative of the cost function for each training input vector and then sum over all input vectors as shown in eq.\ref{eq:gradient}.  First, let us consider the derivative with respect to the weights between output layer and hidden layer.  In this case ($l$=2), we can simply calculate the derivative of $E$ as 

\begin{eqnarray}
 \frac{\partial E_n}{\partial w^{(2)}_{ij}}&=&\frac{\partial E_n}{\partial y_{i,n}}\frac{\partial y_{i,n}}{\partial w^{(2)}_{ij}} \notag\\
 &=&(y_{i,n}-d_{i,n}) t_{j} \notag \\
 &=& (y_{i,n}-d_{i,n}) \phi(s_{j}).
\label{eq:derivative1}
\end{eqnarray}

Next, we calculate the derivative of $E$ with respect to the weights between the hidden layer and the input layer.  In this case ($l$=1), the derivative of $E$ is 

\begin{eqnarray}
 \frac{\partial E_n}{\partial w^{(1)}_{ij}}&=&\frac{\partial E_n}{\partial s_{i}}\frac{\partial s_{i}}{\partial w^{(1)}_{ij}} \notag \\
 &=& \left(\sum_{p=1}^{m}  \frac{\partial E_n}{\partial y_{p}}  \frac{\partial y_{p}}{\partial s_{i}} \right)x_{j} \notag \\
 &=& \left(\sum_{p=1}^{m} (y_{p,n}-d_{p,n})w^{(2)}_{pi}\phi'(s_{i}) \right)x_{j}
\label{eq:derivative2}
\end{eqnarray}

In the second line, we use the chain rule for derivative because $E$ depends on the activated neuron $\phi(s)$ in the hidden layer only through the output neuron $y$. Here, $\phi '$ denotes the derivative of the activation function with respect to $s$. Using eqs.(\ref{eq:gradient}), (\ref{eq:derivative1}) and (\ref{eq:derivative2}), we can iterate on the gradient descent until the outputs obtained by the ANN converge to the desired outputs (minimum of the cost function).  

The back-propagation algorithm can be summarised as follows:

\begin{enumerate}
 \item Starting with random weights, compute the output of the ANN using eq.(\ref{eq:hidden}) and eq.(\ref{eq:output}) for all input vectors in the training set ({\it forward propagation})\\
 
 \item Compute the cost function \\
 
 \item Compute the derivative of the cost function with respect to the weights between output layer and hidden layer with eq.(\ref{eq:derivative1}) and then the derivative with respect to the weights between input layer and hidden layer with eq.(\ref{eq:derivative2}) ({\it back-propagation}).\\
 
 \item Update the weights with eq.(\ref{eq:gradient}).\\
 
 \item Go back to ({\rm i}) and iterate until the cost function converges to a minimum.
 
 \end{enumerate}

\subsection{Training data sets and architecture of the ANN in our work}
In our case, we prepared 70 training data sets.  Each set consist of the 21cm PS $P(k)$ obtained with 21cmFast as input data and the corresponding EoR parameters used in the simulation, $\theta ~(=R_{\rm mfp}, T_{\rm vir}, \zeta)$ as output data.  The architecture of the ANN is the following: ({\it i}) 14 neurons in the input layer,  ({\it ii}) 14 neurons in the hidden layer,  ({\it iii}) 3 neurons in the output layer.  As we mentioned in section~\ref{sec:21cmps}, we use 14~bins for the 21cm PS, and we use 3 EoR parameters.  This is why the number of neurons in the input layer and in the output layers are 14 and 3, respectively.  Note that the number of redshifts used to train will always match the number of redshifts being fit.

To analyse in details how the ANN performs we look at partial cost functions: we use the normalised root mean square error (RMSE) defined by

\begin{equation}
\rm RMSE=\sqrt{\frac{1}{{\it N_{\rm train}}}\sum_{{\it i}=1}^{{\it N}_{\rm train}}{\it X^{2}}}
\end{equation}
Here, $X=(\theta_{\rm ANN}-\theta_{\rm data})/\theta_{\rm data}$  ( with $\theta$ one of $R_{\rm mfp}, T_{\rm vir}$ or, $\zeta)$.  $\theta_{\rm ANN}$ and $\theta_{\rm data}$ are the EoR parameters evaluated by the ANN and from the training data, respectively.

\begin{figure}
\centering
\includegraphics[width=1.0\hsize]{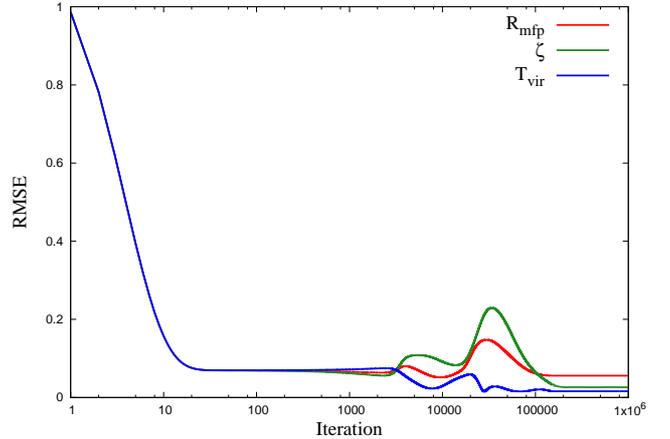}
\caption{Training error (RMSE, see main text for definiton) for $R_{\rm mfp}$ (red), $\zeta$ (green) and $T_{\rm vir}$ (blue) as functions of the number of iterations for the learning process, for a network with 14 neurons.}
\label{fig:fig2}
\end{figure}

\begin{figure}
\centering
\includegraphics[width=1.0\hsize]{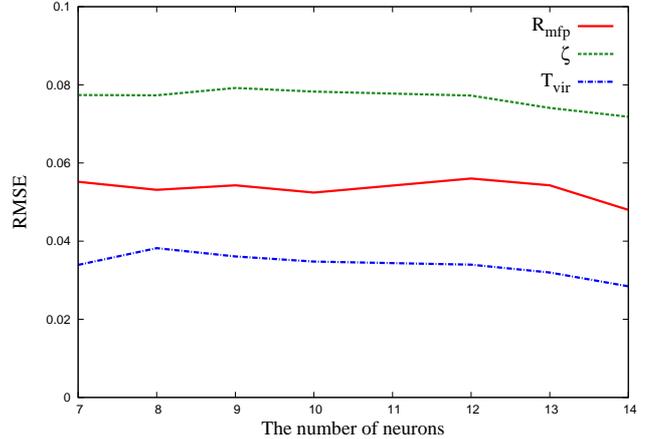}
\caption{Training error (RMSE, see main text for definiton) for  $R_{\rm mfp}$ (solid line), $\zeta$ (dashed line) and $T_{\rm vir}$ (dot-dashed line) as functions of the number of neurons for $10^{5}$ iterations in the learning process. }
\label{fig:fig3}
\end{figure}

First, we study how the RMSE depends on the number of iteration for each EoR parameter. The result is plotted in Fig.\ref{fig:fig2}.  In this figure, we fix the number of neurons in the hidden layer at 14 and perform calculation with $10^{6}$ iterations.  After initial fluctuations, the RMSE decreases and converges for $>100000$ iterations.  Next, we show how the RMSE depends on the number of neurons in the hidden layer after $10^5$ iterations.  The RMSE is only weakly dependent on the number of neurons in the explored range. In the  following section, we perform $10^{5}$ iterations for the back-propagation algorithm with 14 neurons in the hidden layer unless specifically mentioned.

\section{Result}\label{sec:result}
We apply the trained ANN to 54 test datasets.  As for the training datasets, these test datasets consist of the 21cm PS and the corresponding  EoR parameters. We use the 21cm PS data as input for the ANN and obtain estimated values for the EoR parameters.  Then, we compare the estimated values with the true values that were used to compute the PS in the 21cmFast simulation, and evaluate how the ANN performs.

\begin{figure}
\centering
\includegraphics[width=1.0\hsize]{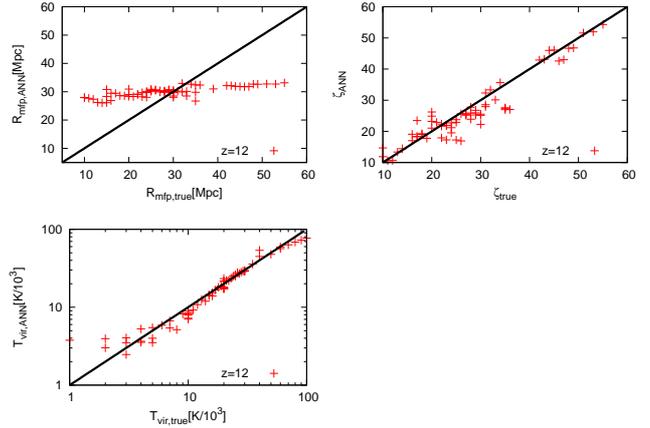}
\caption{The EoR model parameter values computed by the ANN  from the PS against the values used in the simulation at $z$=12.  Note that the result for the Virial temperature is plotted in log scale. It is the same for the following figures.}
\label{fig:fig4}
\end{figure}

\begin{figure}
\centering
\includegraphics[width=1.0\hsize]{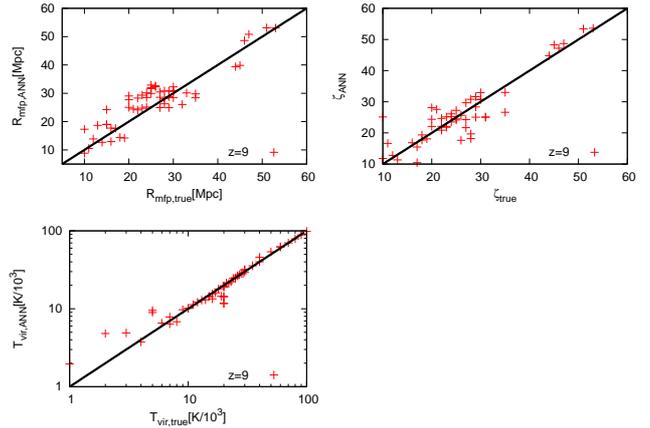}
\caption{The EoR model parameter values computed by the ANN  from the PS against the values used in the simulation at $z$=9.}
\label{fig:fig5}
\end{figure}

First, we apply the ANN to the 21cm PS at a single redshift without including the effect thermal noise or sample variance.  In Figs.\ref{fig:fig4} and \ref{fig:fig5}, we plot the EoR parameters estimated with the ANN from the 21cm PS data against the true values. In Fig.\ref{fig:fig4}, we show the result at $z=12$.  As we can see, $\zeta$ and $T_{\rm vir}$ estimated with the ANN are in relatively good agreement with the true values.  On the other hand, for $R_{\rm mfp}$, the correct values are not recovered. This is because  the 21cm PS is not sensitive at $z=12$ to $R_{\rm mfp}$.  On the other hand, the 21cm PS becomes sensitive to $R_{\rm mfp}$ at lower redshift when reionization advances much more (since $R_{\rm mfp}$ physically expresses the maximum bubble size, the effect of $R_{\rm mfp}$ on the 21cm PS is remarkable mostly at lower redshift).  Thus the cost function is insensitive to changes in $R_{\rm mfp}$ values;  the learning process of the ANN is incomplete and systematic deviations remain. Note that the reason why the estimated value of $R_{\rm mfp}$ seems to be constant at $\sim 30$ is that $R_{\rm mfp}$=30 occurred more often than other values in the training datasets.

In Fig.\ref{fig:fig5} we show the same plots obtained from the PS at $z=9$.  The agreement between recovered and expected values extends over a larger range  than at $z$=12, in particular for $R_{\rm mfp}$.  Indeed the 21cm PS at $z=9$ is sensitive to $R_{\rm mfp}$ and thus the ANN learning process works better.

In the previous two cases, we considered the 21cm PS without any source of noise. We will now consider both the contribution of thermal noise and sample variance. We model the thermal noise as a gaussian random field characterized by its power spectrum. In an annulus in Fourier space with $N_{c}$ cells, the thermal noise power spectrum is  \citep[e.g.][]{2005ApJ...619..678M,2006ApJ...653..815M, 2015MNRAS.451.4785Y}:
\begin{equation}
\delta P^{21}_{\Delta T} (k, \theta) \approx \sqrt{\frac{1}{N_c}}\,
\frac{ A_e \, x^2 \, y }{ \lambda^2 \, B^2} \,
{\it C}^N(k, \theta),
\label{eq:dP}
\end{equation}
where $\theta$ is the angle between {\bf k} and line of sight, $A_{e}$ is the effective area of antenna, $\lambda$ is the observed wavelength and $x, y$ are factors converting $uv$ space units into comoving wavenumber units and are determined by cosmology.  $C^{N}$ is expressed by 

\begin{equation}
C^N(k,\theta) = \left(\frac{\lambda^2 \, B \, T_{\rm sys}}{
A_e}\right)^2 \, \frac{1}{B \, t_{\bf k}},
\label{eq:CN}
\end{equation}
where $T_{\rm sys}$ is the total system temperature, $B$ is the bandwidth and $t_{\bf k}$ is the effective observing time.  The spherically averaged noise can be obtained by summing over $\theta$ in a shell with radius $k$ shown in eq.\ref{eq:noise_average}.

\begin{equation}
\delta P_{\Delta T}^{21}(k)=\left\{\sum_{\theta}\left[\frac{1}{\delta P^{21}_{\Delta T} (k, \theta)}\right]^{2}\right\}^{-1/2}.
\label{eq:noise_average}
\end{equation}

We recommend to read McQuinn et al. (2006).  In our case, we assume SKA specification \cite{SKA}. We estimate the sample variance from 10 simulations using different realizations of the initial conditions.

In the general case, we would include the noise at the level of the visibilities and compute the resulting noisy power spectrum. Assuming that the noise is a random gaussian field, we simply generate a noisy power spectrum by using the following formula:

\begin{equation}
P^{ij}_N(k)=P^i(k) + N^j(k)
\end{equation}

where $P^{ij}_N(k)$ is the noisy power spectrum, $P^j(k)$ is the power spectrum produced by the simulation for a given set of parameters generically labeled by the index j, and $N^i(k)$ is \textit{a random draw} from a Gaussian probability distribution with variance equal to the thermal noise power spectrum or the sample variance. A different and independent draw is required for each value of $k$.
For the learning sample we include only the sample variance. For each parameter set (70 possible values for $i$), we generate 50 realizations of the noise (labeled by $j=1 \dots 50$).  This means that we use 3500 different PS for the learning sample. This is necessary as we are not aware of a standard technique for directly including an uncertainty in the inputs of an ANN. For the test datasets, we add both thermal noise and sample variance to the 21cm PS, using the same procedure.

%
%


\begin{figure}
\centering
\includegraphics[width=1.0\hsize]{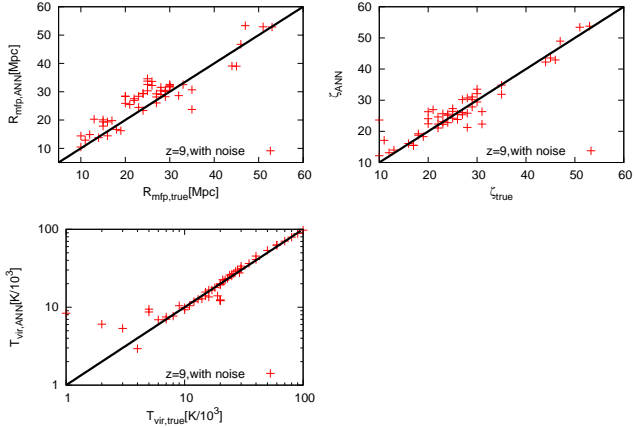}
\caption{The EoR model parameter values computed by the ANN  from the PS against the values used in the simulation at $z$=12.  In this case, we include both thermal noise and sample variance. }
\label{fig:fig6}
\end{figure}
In Fig.\ref{fig:fig6}, we show the EoR parameters found by the ANN as functions of the values used in the simulations, using the 21cm signal PS at $z=9$, including both thermal noise and sample variance. The difference between Fig.\ref{fig:fig5} and Fig.\ref{fig:fig6} is not obvious at a glance.  In order to quantify the difference, we compute the {\it mean chi-square value}, $\chi^{2}$:

\begin{equation}
\chi^{2}=\sqrt{\frac{1}{N_{\rm test}}\sum_{i=1}^{N_{\rm test}} \frac{(\theta_{\rm ANN}-\theta_{\rm true})^{2}}{\theta_{\rm true}^{2}}}
\end{equation}
$\theta_{\rm ANN}$ is the  $R_{\rm mfp}$, $T_{\rm vir}$, $\zeta$ reconstructed by the ANN and $\theta_{\rm true}$ is the value of the corresponding parameter used in the simulation.  In table.\ref{table:chi}, we compare the $\chi^{2}$  values for each of the parameters, with and without noise.  As we can see, the $\chi^{2}$ values for the 21cm PS with noises are worse than those without noises.  Noise alters the efficiency of the learning process for the ANN.

\begin{table*}
\centering
\label{my-label}
\begin{tabular}{llllll}
   & $\chi^{2}_{\rm wo/noise}$ & $\chi^{2}_{\rm w/noise}$ & $\chi^{2}_{\rm w/noise,zevolution}$ & $\chi^{2}_{\rm w/noise, reduced}$ \\ \hline
$R_{\rm mfp}$& 0.228    & 0.258    &  0.172 & 0.262\\ \hline
$\zeta$ & 0.271   & 0.288    & 0.168&  0.290 \\ \hline
$\log(T_{\rm vir})$&   0.027   & 0.038    & 0.019& 0.029\\ \hline
 \end{tabular}
 \caption{The mean chi-square values for the EoR model parameter values computed by the ANN  from the PS as functions of the values used in the simulations.  We show mean chi-square values for four types of data.  The first column of this table is the chi-square values when we do not include any noise, in second column we take thermal noise and sample variance into account. We use the 21cm PS at $z=9$ for both cases. The third column is when we take redshift evolution($z=9,10,11)$ into account and the fourth column is when we take both thermal noise and sample variance into account and we reduce the number of training data samples.}
\label{table:chi}
\end{table*}

\begin{figure}
\centering
\includegraphics[width=1.0\hsize]{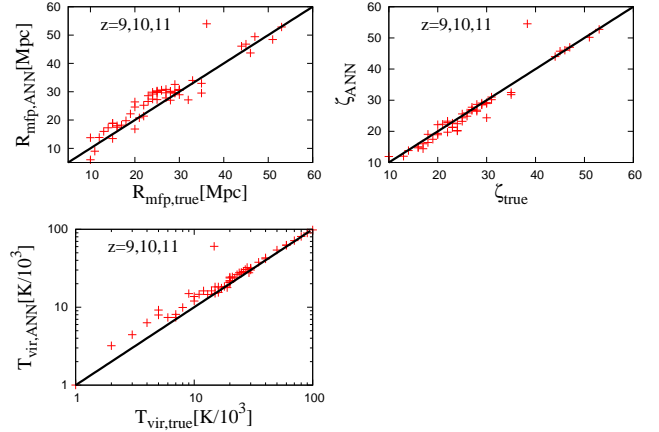}
\caption{Comparing the EoR model parameter values computed by the ANN   against the values used in the simulation using the PS at $z=9, 10$, and $11$ as input data and including both thermal noise and sample variance. }
\label{fig:fig7}
\end{figure}

Until now we have used as input data for the ANN the 21cm PS at a single redshift. We now expand our input datas to take the redshift evolution of the 21cm PS into account.  We use the 21cm PS, which includes thermal noise and sample variance, at three redshifts $z=9,10$ and $11$ as input datas for the training and test datasets.  Thus each set as 14$\times$3 input values and the  number of neurons in the input layer is also 14$\times$3=42. We also increase the number of neurons from 14 to 42 in the hidden layer. Note that the number of training datasets is unchanged.  In Fig.\ref{fig:fig7}, we show parameter values computed by the ANN as functions of the values used in the simulations when the ANN is using the 21cm PS at three different redshifts. Compared with the single redshift case, it seems that the scatter is smaller.  We quantify this in table.\ref{table:chi}, and check that the scatter is indeed smaller for all three parameters. The reason why using the 21cm PS at multiple redshifts improves the accuracy of the ANN results is simply that we increase the information the ANN has to work with.  At each redshift, the dependence and sensitivity of the 21cm PS on the parameters is different.  Therefore, the trained network can learn much more if we consider the redshift evolution of the 21cm PS.   We also notice that the chi-square values obtained by using multiple redshifts and including noise are smaller than those for a single redshift without  any noise.  

\begin{figure}
\centering
\includegraphics[width=1.0\hsize]{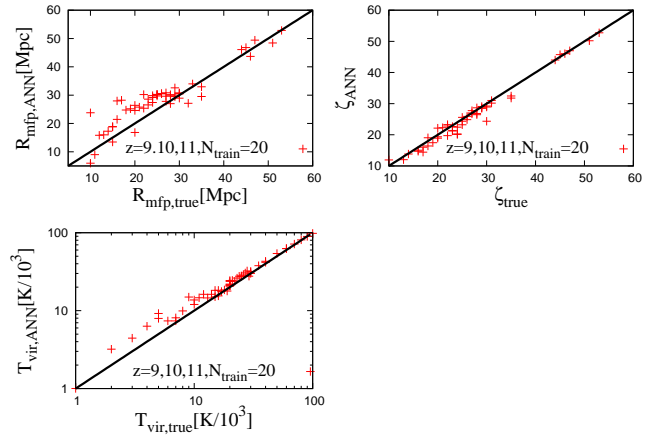}
\caption{The EoR model parameter values computed by the ANN  for the PS against the values used in the simulation at $z$=9, using only 20 training datasets during the learning process. }
\label{fig:fig8}
\end{figure}

We next check the influence of the size of the training sample.  If we reduce the number of training datasets, we expect that the learning efficiency of the ANN will decrease.  To check this, we train the ANN with a subsample of the training datasets ($N_{\rm train}$=20). We show the results in Fig.\ref{fig:fig8}.  As expected, we can see that the scatter of reconstructed parameter values is larger when using fewer training datasets. This is quantitified in table.\ref{table:chi}.  In particular, the scatter of $R_{\rm mfp}$ increases more than for the other parameters.  This is connected to the fact already established that the PS is not very sensitive to $R_{\rm mfp}$.  Thus the size of the learning sample is crucial to  make accurate EoR parameter predictions from observed data.  

\section{Summary \& Discussion}\label{sec:summary}
In this paper, we applied artificial neural network~(ANN) to estimate EoR parameters from the 21cm power spectrum~(PS) .  We used 21cmFAST to produce 21cm PS for different values of the following parameters: $R_{\rm mfp}, \zeta$ and $T_{\rm vir}$.  We ran $70$ simulations, that provided us with $70$ training datasets made of the 21cm PS and the corresponding EoR parameter values.  With these datasets, we trained the ANN, which consists of an input layer, one hidden layer and an output layer. We achieved the training by applying the back-propagation algorithm.  

In order to estimate the quality of the learning process of the ANN, we first show the root mean square error~(RMSE) for the training datasets depends on the number of iterations of the back-propagation algorithm and on the number of neurons in the hidden layer.  We found that the back-propagation algorithm converges as the number of iteration increases and weakly depends on the number of neurons in the hidden layer. 

 We next applied the trained network to test datasets made of the 21cm PS at a single redshift including or not thermal noise and sample variance.  As expected, we found that the chi-square values for the ANN-recovered values of the parameters as functions of the true (simulation) values are larger when including thermal noise and sample variance. We also found that the $\zeta$ and $T_{\rm vir}$ values obtained by the ANN were a better match to the true values than for $R_{\rm mfp}$ due the different sensitivity of the 21cm PS to the different parameters.  
 
 Then we increased the information in the inputs given to the ANN by including the PS at three different redshifts.  We found that  the accuracy of parameters computed by the ANN was improved compared with those obtained based on the PS at a single redshift.  This is simply that the neural network takes advantage in its learning process of the  fact that the input hold more information and that the 21cm PS is sensitive to different parameters at different redshifts.  
 
One of the remarkable points of our study is that we can recover the EoR parameters from the power spectrum using the ANN with good accuracy although we only use 70 training datasets.  This number is quite small compared with that required for MCMC (e.g. $10^{5}$ realizations of the PS are used in 21CMMC \cite{2015MNRAS.449.4246G}). In table.\ref{table:chi}, we show the mean chi-square value for each parameter calculated by the ANN. To enable a quantitative comparison with 21CMMC, we perform a parameter estimation with the ANN for HERA and SKA observations and compare to the $1\sigma$ uncertainties obtained with 21CMMC (for HERA, we actually use the noise data provided in 21CMMC package). We show the result in table. \ref{table:chi2}. Using a single redshift, the accuracy of our reconstruction is of the same order as that of 21CMMC, although somewhat worse. Remember however that it does improve when using several redshifts, as does the accuracy of 21CMMC.  However, note that the MCMC method finds the local shape of the likelihood space, while the ANN method only finds the average error over a swath of parameter space. So, while the ANN method requires a far sparser exploration of the parameter space, 21CMMC does provide more detailed informations. 
In addition, we need to mention the effect of the foregrounds. The foregrounds wedge may hinder the use of part of the power spectrum. Thus taking this into account would likely affect the constraints on the $R_{\rm mfp}$ because it depends heavily on the available range of $k$-space. We defer an ANN analysis incorporating the effect of foregrounds to a future work.  Both the training process for the ANN and the parameter estimation from the observed data using the trained ANN have a very small computing cost compared to that required for producing training datasets.

In this work, we used the 21cm PS generated by 21cmFAST.  However, it is worth applying the ANN to other EoR modelling methods including full simulations in order to check its ability to reconstruct the astrophysical parameters.   In our model, we focused on the EoR and used the 21cm PS.  Thus, a possible development is to expand the model beyond the EoR and include the Cosmic Dawn.  During the Cosmic Dawn, X-ray heating of the IGM plays an important role. Thus, it is important to incorporate X-ray heating models and study their impact on the 21 cm \cite{2010A&A...523A...4B, 2013MNRAS.431..621M,2016MNRAS.460.4320E,2017MNRAS.464.3498F}.  Recently, \cite{2016arXiv160902312C} explored the behaviour of the 21cm global signal in a large parameter space from $z=40$ to $z=6$.  We believe it is worth applying the ANN method using the 21cm global signal as an input and try to reconstruct parameter characteristic of the Cosmic Dawn and EoR both.

Another possible application of ANNs is to use the 21cm tomography as an input. Then the learning sample is made of light cone data from 21cm  simulations instead of the 21cm PS.  Although the 21cm PS is a powerful tool, the $21$ cm signal is not gaussian, and thus the PS does not contain all information. The 21cm light cone data on the other hand holds all the information that will ever be accessible trough observations. It also contains irrelevant information that the ANN will have to filter out: the random phases of the initial density field.

If we consider a large survey field of view and depth, that is a large box size for the learning sample simulation, we will need to treat large amount of data with the ANN.  Machine learning method have been developed to treat such big data.  One of the methods that have received attention is deep learning \citep[e.g.][]{2014arXiv1404.7828S,LeCun}.  In our case, we only used 1 hidden layer in the ANN.  A deep learning neural network consists of one input layer, multiple hidden layers and one output layer. Deep learning has been applied successfully to image recognition.  Therefore, one of the our future works will be to apply deep learning to 21 cm light cones from full simulations data.

\begin{table}
\centering
\label{my-label}
\begin{tabular}{lllll}
 & $\chi^{2}_{\rm SKA}$ &$\chi^{2}_{\rm HERA}$ &1$\sigma_{\rm SKA}$ & 1$\sigma_{\rm HERA}$ \\ \hline
$R_{\rm mfp}$ & 0.258& 0.278& 0.178 & 0.184\\ \hline
$\zeta$ &0.288 &  0.354 & 0.167 & 0.220 \\ \hline
$\log(T_{\rm vir})$ & 0.038 & 0.040 & 0.024 &0.033 \\ \hline
 \end{tabular}
 \caption{We compare the mean chi-square value obtained by the ANN with the 1 $\sigma$ fractional error obtained by 21CMMC for HERA and SKA observations at $z=9$.  For comparison, we adopt the noise used in 21CMMC for HERA.} 
\label{table:chi2}
\end{table}

\section*{Appendix}
Here, we show the back propagation algorithm in the case of multiple hidden layers.  By analogy with eq.(\ref{eq:derivative2}), we can express the derivative of $E_{n}$ with respect to the weight in the $l$-th layer, $w_{ij}^{(l)}$ $(l=1,2,...L-1)$ as

\begin{equation}
 \frac{\partial E_n}{\partial w^{(l)}_{ij}}=\frac{\partial E_n}{\partial s^{(l)}_{i}}\frac{\partial s^{(l)}_{i}}{\partial w^{(l)}_{ij}}, 
\label{eq:derivativel}
\end{equation}
where $s_{i}^{(l)}=\sum w_{ij}^{(l)}t^{(l-1)}_{j}$.  Since the changes in $s_{i}^{(l)}$ are transmitted to $E_{n}$ through each $s_{k}^{(l+1)}$ in the ($l$+1)-th layer, the derivative of $E_{n}$ with respect to $s_{i}^{(l)}$ can be expressed as

\begin{equation}
\frac{\partial E_n}{\partial s^{(l)}_{i}}=\sum_{k}\frac{\partial E_n}{\partial s^{(l+1)}_{k}}\frac{\partial s_{k}^{(l+1)}}{\partial s_{i}^{(l)}},
\label{eq:derivativel2}
\end{equation}

Remember that $s_{k}^{(l+1)}$ can be expressed with the activation function $\phi$ as $s_{k}^{(l+1)}=\sum w_{kj}^{(l+1)}t^{(l)}_{j}=\sum w_{kj}^{(l+1)}\phi(s_{j}^{(l)})$,

\begin{equation}
 \frac{\partial s_{k}^{(l+1)}}{\partial s_{i}^{(l)}}=w_{ki}^{(l+1)}\phi'(s_{i}^{(l)}).
\end{equation}

Here, we define $\delta_{i}^{(l)}\equiv \partial E_n/\partial s^{(l)}_{i}$, then we can re-write eq.(\ref{eq:derivativel2}) as 

\begin{equation}
\delta_{i}^{(l)}=\sum_{k}\delta_{k}^{(l+1)}w_{ki}^{(l+1)}\phi'(s_{i}^{(l)}).
\label{eq:delta}
\end{equation}
Combining eq.(\ref{eq:delta}) with $\partial s^{(l)}_{i}/\partial w^{(l)}_{ij}=t_{j}^{(l-1)}$, the eq.(\ref{eq:derivativel}) can be re-written simply as 

\begin{equation}
 \frac{\partial E_n}{\partial w^{(l)}_{ij}}=\delta_{i}^{(l)}t_{j}^{(l-1)}.
\end{equation}

This form tells us that we can easily obtain the derivative of the cost function with respect to $w_{ij}^{(l)}$, which connects the neuron $j$ in the ($l$-1) th layer to the neuron $i$ in the $l$ th layer, as the product of $\delta_{i}^{(l)}$ and $t_{j}^{(l-1)}$.  As shown in eq.(\ref{eq:delta}), we start to calculate $\delta_{i}^{(l)}$ from the output layer ($l$=L) to the input layer. This is why this algorithm is called ``{\it back propagation}''.  If we use eq.(\ref{eq:cost}) as the cost function, we easily calculate $\delta_{i}^{(L)}=y_{i}-d_{i}$.

\section*{Acknowledgement}
This work is benefited from a grant from the French ANR funded project ORAGE (ANR-14- CE33-0016).  We thank to G.Mellema, A. Fialkov, S. Majumdar, S.Giri and K. Hasegawa for their useful comments and thank to S. Yoshiura  for providing the thermal noise data.


\label{lastpage}


\begin{thebibliography}{99}

\bibitem[Abel et al. 2002]{2002Sci...295...93A} Abel, T., Bryan, G.~L., \& Norman, M.~L.\ 2002, Science, 295, 93 
\bibitem[Agarwal et al. 2012]{2012MNRAS.424.1409A} Agarwal, S., Abdalla, F.~B., Feldman, H.~A., Lahav, O., \& Thomas, S.~A.\ 2012, \mnras, 424, 1409 
\bibitem[Agarwal et al. 2014]{2014MNRAS.439.2102A} Agarwal, S., Abdalla, F.~B., Feldman, H.~A., Lahav, O., \& Thomas, S.~A.\ 2014, \mnras, 439, 2102 

\bibitem[Ali et al. 2015]{2015ApJ...809...61A} Ali, Z.~S., Parsons, A.~R., Zheng, H., et al.\ 2015, \apj, 809, 61 

\bibitem[Baek et al. 2010]{2010A&A...523A...4B} Baek, S., Semelin, B., Di Matteo, P., Revaz, Y., \& Combes, F.\ 2010, \aap, 523, A4 

\bibitem[Ball \& Brunner 2010]{2010IJMPD..19.1049B} Ball, N.~M., \& Brunner, R.~J.\ 2010, International Journal of Modern Physics D, 19, 1049 
\bibitem[Banerji et al. 2010]{2010MNRAS.406..342B} Banerji, M., Lahav, O., Lintott, C.~J., et al.\ 2010, \mnras, 406, 342 
\bibitem[Barkana \& Loeb 2008]{2008MNRAS.384.1069B} Barkana, R., \& Loeb, A.\ 2008, \mnras, 384, 1069 

\bibitem[Beardsley et al. 2016]{2016ApJ...833..102B} Beardsley, A.~P., Hazelton, B.~J., Sullivan, I.~S., et al.\ 2016, \apj, 833, 102 
\bibitem[Bernardi et al. 2016]{2016MNRAS.461.2847B} Bernardi, G., Zwart, J.~T.~L., Price, D., et al.\ 2016, \mnras, 461, 2847 
\bibitem[Bloom \& Richards 2012]{2012amld.book...89B} Bloom, J.~S., \& Richards, J.~W.\ 2012, Advances in Machine Learning and Data Mining for Astronomy, 89 
\bibitem[Bromm et al. 2002]{2002ApJ...564...23B} Bromm, V., Coppi, P.~S., \& Larson, R.~B.\ 2002, \apj, 564, 23 
\bibitem[Bromm 2013]{2013RPPh...76k2901B} Bromm, V.\ 2013, Reports on Progress in Physics, 76, 112901 




\bibitem[Ciardi et al. 2000]{2000MNRAS.314..611C} Ciardi, B., Ferrara, A., Governato, F., \& Jenkins, A.\ 2000, \mnras, 314, 611 

\bibitem[Ciardi et al. 2006]{2006MNRAS.366..689C} Ciardi, B., Scannapieco, E., Stoehr, F., et al.\ 2006, \mnras, 366, 689 

\bibitem[Cohen et al. 2016]{2016arXiv160902312C} Cohen, A., Fialkov, A., Barkana, R., \& Lotem, M.\ 2016, arXiv:1609.02312 
\bibitem[Collister \& Lahav 2004]{2004PASP..116..345C} Collister, A.~A., \& Lahav, O.\ 2004, \pasp, 116, 345 

\bibitem[Cybenko. 1989]{Cybenko}
Cybenko.G,
{\it Mathematics of control. Signals and Systems}, Vol. 2 pp. 303-314,1989


\bibitem[Dewdney. 2013]{SKA}
Dewdney, P., Turner, W., Millenaar, R., McCool, R., Lazio, J.
\& Cornwell, T., 2013, ”SKA1 System Baseline Design”,
Document number SKA-TEL-SKO-DD-001 Revision 1

\bibitem[Ewall-Wice et al. 2016]{2016MNRAS.460.4320E} Ewall-Wice, A., Dillon, J.~S., Hewitt, J.~N., et al.\ 2016, \mnras, 460, 4320 

%
%
\bibitem[Fan et al. 2006]{Fan:2006dp}
  Fan. H. X, Carilli. L. C and Keating. B.~G.
  Ann.\ Rev.\ Astron.\ Astrophys.\  {\bf 44} (2006) 415
  [astro-ph/0602375].
%

\bibitem[Furlanetto et al. 2004]{2004ApJ...613....1F} Furlanetto, S.~R., Zaldarriaga, M., \& Hernquist, L.\ 2004, \apj, 613, 1 

\bibitem[Furlanetto et al. 2006]{fur}
Furlanetto. S, Oh. P. S and Briggs. F,
  Phys.\ Rept.\  {\bf 433} (2006) 181
  [astro-ph/0608032].
%
 \bibitem[DeBoer et al. 2016]{2016arXiv160607473D} DeBoer, D.~R., Parsons, A.~R., Aguirre, J.~E., et al.\ 2016, arXiv:1606.07473 DeBoer, D.~R., Parsons, A.~R., Aguirre, J.~E., et al.\ 2016, arXiv:1606.07473 

\bibitem[Fialkov et al. 2013]{2013MNRAS.432.2909F} Fialkov, A., Barkana, R., Visbal, E., Tseliakhovich, D., \& Hirata, C.~M.\ 2013, \mnras, 432, 2909 
\bibitem[Fialkov et al. 2017]{2017MNRAS.464.3498F} Fialkov, A., Cohen, A., Barkana, R., \& Silk, J.\ 2017, \mnras, 464, 3498 
\bibitem[Folkes et al. 1996]{1996MNRAS.283..651F} Folkes, S.~R., Lahav, O., \& Maddox, S.~J.\ 1996, \mnras, 283, 651 

\bibitem[Furlanetto et al. 2004]{2004ApJ...613....1F} Furlanetto, S.~R., Zaldarriaga, M., \& Hernquist, L.\ 2004, \apj, 613, 1 

%
\bibitem[Greig \& Mesinger 2015]{2015MNRAS.449.4246G} Greig, B., \& Mesinger, A.\ 2015, \mnras, 449, 4246
\bibitem[Greig et al. 2015]{2015arXiv150903312G} Greig, B., Mesinger, A., \& Koopmans, L.~V.~E.\ 2015, arXiv:1509.03312 
\bibitem[Greig et al. 2016]{2016MNRAS.455.4295G} Greig, B., Mesinger, A., \& Pober, J.~C.\ 2016, \mnras, 455, 4295 
\bibitem[Gnedin et al. 2008]{2008ApJ...672..765G} Gnedin, N.~Y., Kravtsov, A.~V., \& Chen, H.-W.\ 2008, \apj, 672, 765-775 


\bibitem[Iliev et al. 2006]{2006MNRAS.369.1625I} Iliev, I.~T., Mellema, G., Pen, U.-L., et al.\ 2006, \mnras, 369, 1625 

\bibitem[Iliev et al. 2012]{2012MNRAS.423.2222I} Iliev, I.~T., Mellema, G., Shapiro, P.~R., et al.\ 2012, \mnras, 423, 2222 
\bibitem[Iliev et al. 2014]{2014MNRAS.439..725I} Iliev, I.~T., Mellema, G., Ahn, K., et al.\ 2014, \mnras, 439, 725 

\bibitem[Harker et al. 2009]{2009MNRAS.393.1449H} Harker, G.~J.~A., Zaroubi, S., Thomas, R.~M., et al.\ 2009, \mnras, 393, 1449 
\bibitem[Hasegawa \& Semelin 2013]{2013MNRAS.428..154H} Hasegawa, K., \& Semelin, B.\ 2013, \mnras, 428, 154 

\bibitem[Hornik et al. 1989]{Hornik} Hornik, K., Stinchcombe,M., White, H., {\it Neural Networks} Vol.2 pp359-366, 1989
%

\bibitem[Jacobs et al.2015]{2015ApJ...801...51J} Jacobs, D.~C., Pober, J.~C., Parsons, A.~R., et al.\ 2015, \apj, 801, 51 
\bibitem[Jensen et al. 2016]{2016ApJ...827....5J} Jensen, H., Zackrisson, E., Pelckmans, K., et al.\ 2016, \apj, 827, 5 



\bibitem[Kamdar et al. 2016a]{2016MNRAS.455..642K} Kamdar, H.~M., Turk, M.~J., \& Brunner, R.~J.\ 2016, \mnras, 455, 642 
\bibitem[Kamdar et al. 2016b]{2016MNRAS.457.1162K} Kamdar, H.~M., Turk, M.~J., \& Brunner, R.~J.\ 2016, \mnras, 457, 1162 
\bibitem[Kim et al. 2015]{2015CQGra..32x5002K} Kim, K., Harry, I.~W., Hodge, K.~A., et al.\ 2015, Classical and Quantum Gravity, 32, 245002 

 \bibitem[Konno et al. 2014]{2014ApJ...797...16K} Konno, A., Ouchi, M., 
Ono, Y., et al.\ 2014, ApJ, 797, 16 

\bibitem[Kubota et al. 2016]{2016PASJ...68...61K} Kubota, K., Yoshiura, S., Shimabukuro, H., \& Takahashi, K.\ 2016, \pasj, 68, 61 

\bibitem[Lahav et al. 1996]{1996MNRAS.283..207L} Lahav, O., Naim, A., Sodr{\'e}, L., Jr., \& Storrie-Lombardi, M.~C.\ 1996, \mnras, 283, 207 

\bibitem[LeCun et al. 2015]{LeCun} LeCun, Y, Bengio, Y \& Hinton, G. 2015  {\it Nature}, 521,7553,pp436-444

\bibitem[Liu \& Parsons 2016]{2016MNRAS.457.1864L} Liu, A., \& Parsons, A.~R.\ 2016, \mnras, 457, 1864 

 \bibitem[McQuinn et al. 2006]{2006ApJ...653..815M}
McQuinn, M., Zahn, O., Zaldarriaga, M., Hernquist, L., \& Furlanetto, S.~R.\ 2006, ApJ, 653, 815

\bibitem[McQuinn et al. 2007]{2007MNRAS.377.1043M}  McQuinn, M., Lidz, A., Zahn, O., et al.\ 2007, \mnras, 377, 1043 

\bibitem[McQuinn et al. 2011]{2011ApJ...743...82M} McQuinn, M., Oh, S.~P., \& Faucher-Gigu{\`e}re, C.-A.\ 2011, \apj, 743, 82 

  \bibitem[Mellema et al. 2013]{2013ExA....36..235M} Mellema, G., Koopmans, L.~V.~E., Abdalla, F.~A., et al.\ 2013, Experimental Astronomy, 36, 235 
%
%
\bibitem[Mesinger \& Furnaletto 2007]{Mesinger:2007pd} 
  Mesinger. A and Furlanetto. S,
  arXiv:0704.0946 [astro-ph].
%
  \bibitem[Mesinger et al. 2011]{2011MNRAS.411..955M} 
Mesinger. A, Furlanetto. S, Cen. R,2011, MNRAS, 411, 955 
\bibitem[Mesinger et al. 2013]{2013MNRAS.431..621M} Mesinger, A., Ferrara, A., \& Spiegel, D.~S.\ 2013, \mnras, 431, 621 

\bibitem[Mesinger et al. 2014]{2014MNRAS.439.3262M} Mesinger, A., Ewall-Wice, A., \& Hewitt, J.\ 2014, \mnras, 439, 3262 

\bibitem[Morales 2005]{2005ApJ...619..678M} Morales, M.~F.\ 2005, \apj, 619, 678 
\bibitem[Morii et al. 2016]{2016PASJ...68..104M} Morii, M., Ikeda, S., Tominaga, N., et al.\ 2016, \pasj, 68, 104 
\bibitem[Mukund et al. 2016]{2016arXiv160907259M} Mukund, N., Abraham, S., Kandhasamy, S., Mitra, S., \& Sajeeth Philip, N.\ 2016, arXiv:1609.07259 


\bibitem[Parsons et al. 2014]{2014ApJ...788..106P} Parsons, A.~R., Liu, A., Aguirre, J.~E., et al.\ 2014, \apj, 788, 106 
\bibitem[Patil et al. 2014]{2014MNRAS.443.1113P} Patil, A.~H., Zaroubi, S., Chapman, E., et al.\ 2014, \mnras, 443, 1113 


\bibitem[Planck Collaboration et al. 2016]{2016arXiv160503507P} Planck Collaboration, Adam, R., Aghanim, N., et al.\ 2016, arXiv:1605.03507

\bibitem[Planck Collaboration et al. 2016]{2016A&A...594A..13P} Planck Collaboration, Ade, P.~A.~R., Aghanim, N., et al. \ 2016, \aap, 594, A13 

\bibitem[Pober et al. 2014]{2014ApJ...782...66P} Pober, J.~C., Liu, A., Dillon, J.~S., et al.\ 2014, \apj, 782, 66 

\bibitem[Pober et al. 2015 ]{2015ApJ...809...62P} Pober, J.~C., Ali, Z.~S., Parsons, A.~R., et al.\ 2015, \apj, 809, 62 

 \bibitem[Pritchard \& Loeb 2012]{2012RPPh...75h6901P} Pritchard, J.~R., \& Loeb, A.\ 2012, Reports on Progress in Physics, 75, 086901 

\bibitem[Pritchard et al. 2015]{2015aska.confE..12P} Pritchard, J., 
Ichiki, K., Mesinger, A., et al.\ 2015, Advancing Astrophysics with the 
Square Kilometre Array (AASKA14), 12 
%
\bibitem[Ricotti et al. 2002]{2002ApJ...575...49R} Ricotti, M., Gnedin, N.~Y., \& Shull, J.~M.\ 2002, \apj, 575, 49 

\bibitem[Rottgering 2003]{Rottgering:2003jh}
 Rottgering. H,
  New Astron.\ Rev.\  {\bf 47} (2003) 405
  [astro-ph/0309537].

\bibitem[Rumelhart et al.1986]{1986Natur.323..533R} Rumelhart, D.~E., Hinton, G.~E., \& Williams, R.~J.\ 1986, \nat, 323, 533

\bibitem[Samui \& Samui Pal 2017]{2017NewA...51..169S} Samui, S., \& Samui Pal, S.\ 2017, \na, 51, 169 

\bibitem[Santos et al. 2010]{2010MNRAS.406.2421S} Santos, M.~G., Ferramacho, L., Silva, M.~B., Amblard, A., \& Cooray, A.\ 2010, \mnras, 406, 2421\bibitem[Schmidhuber 2014]{2014arXiv1404.7828S} Schmidhuber, J.\ 2014, arXiv:1404.7828 

\bibitem[Semelin \& Iliev 2015]{2015aska.confE..13S} Semelin, B., \& Iliev, I.\ 2015, Advancing Astrophysics with the Square Kilometre Array (AASKA14), 13 

\bibitem[Shimabukuro et al. 2015]{2015MNRAS.451.4986S} Shimabukuro. H, 
Yoshiura. S, Takahashi. K, Yokoyama. S 
\& Ichiki, K.\ 2015,  Mon.\ Not.\ Roy.\ Astron.\ Soc.\ , 451, 4986 


\bibitem[Shimabukuro et al. 2016a]{2016MNRAS.458.3003S} Shimabukuro, H., Yoshiura, S., Takahashi, K., Yokoyama, S., \& Ichiki, K.\ 2016, \mnras, 458, 3003 

\bibitem[Shimabukuro et al. 2016b]{2016arXiv160800372S} Shimabukuro, H., Yoshiura, S., Takahashi, K., Yokoyama, S., \& Ichiki, K.\ 2016, arXiv:1608.00372 
%
%
\bibitem[Sobacchi \& Mesinger. 2013]{2013MNRAS.432.3340S} Sobacchi, E., \& Mesinger, A.\ 2013,\mnras, 432, 3340
%
%
\bibitem[Sobacchi \& Mesinger. 2014]{2014MNRAS.440.1662S}  Sobacchi, E., \& Mesinger, A.\ 2014, \mnras, 440, 1662 

\bibitem[Tingay et al. 2012]{Tingay:2012ps}
  Tingay. S.~J, Goeke. R, Bowman. J.~D, Emrich. D, Ord. S.~M, Mitchell. D.~A, Morales. M.~F and Booler. T {\it et al.},
  arXiv:1206.6945 [astro-ph.IM].
%
\bibitem[Trac \& Cen 2007]{2007ApJ...671....1T} Trac, H., \& Cen, R.\ 2007, \apj, 671, 1 

\bibitem[Vanzella et al. 2004]{2004A&A...423..761V} Vanzella, E., Cristiani, S., Fontana, A., et al.\ 2004, \aap, 423, 761 


\bibitem[Wise \& Cen 2009]{2009ApJ...693..984W} Wise, J.~H., \& Cen, R.\ 2009, \apj, 693, 984 


\bibitem[Yoshida et al. 2006]{yos}
Yoshida. N, Omukai. K, Hernquist. L, \& T.~Abel \ 2006, ApJ, 652, 6
%
\bibitem[Yoshiura et al. 2015]{2015MNRAS.451.4785Y} Yoshiura.S, 
Shimabukuro.H, Takahashi.K, et al.\ 2015,  Mon.\ Not.\ Roy.\ Astron.\ Soc.\, 451, 4785 
%

\bibitem[Zahn et al. 2007]{2007ApJ...654...12Z} Zahn, O., Lidz, A., McQuinn, M., et al.\ 2007, \apj, 654, 12 
\bibitem[Zahn et al. 2011]{2011MNRAS.414..727Z} Zahn, O., Mesinger, A., McQuinn, M., et al.\ 2011, \mnras, 414, 727 





    
\end{thebibliography}
\end{document}